\documentclass[aps,prd,twocolumn,amsmath,amssymb,showpacs,floatfix,superscriptaddress,nofootinbib]{revtex4-1}

\pdfoutput=1
\usepackage{natbib}
\usepackage{amsmath}
\usepackage{amssymb}
\usepackage{latexsym}
\usepackage{bm}   
\usepackage{graphicx,epstopdf}
\usepackage{epstopdf}
\DeclareGraphicsExtensions{.ps}
\usepackage{dcolumn} 
\usepackage{multirow}
\usepackage{appendix}
\usepackage{footnote}
\usepackage{tabularx,ragged2e,booktabs}

\newcolumntype{C}[1]{>{\Centering}m{#1}}

\newcommand{\refsec}[1]{section~\ref{sec:#1}}

\newcommand{\eqq}[1]{Eq.~(\ref{eq:#1})}

\newcommand{\vx}{\boldsymbol{x}}
\newcommand{\VX}{\boldsymbol{X}}
\newcommand{\vk}{\boldsymbol{k}}
\newcommand{\VK}{\boldsymbol{K}}
\newcommand{\hMpc}{h^{-1}{\rm Mpc}}
\newcommand{\ihMpc}{h\,{\rm Mpc}^{-1}}  
\newcommand{\hGpc}{h^{-1}{\rm Gpc}}

\newcommand{\DD}{_L^{\Delta\Delta}}

\newcommand{\beq}{\begin{equation}}
\newcommand{\eeq}{\end{equation}}
\newcommand{\bea}{\begin{eqnarray}}
\newcommand{\eea}{\end{eqnarray}}
\newcommand{\fig}[1]{Fig.~\ref{fig:#1}} 

\usepackage{color}
\definecolor{ultramarine}{rgb}{0.07, 0.04, 0.56}
\definecolor{cadmiumgreen}{rgb}{0.0, 0.42, 0.24}
\definecolor{indigo(dye)}{rgb}{0.0, 0.25, 0.42}
\usepackage[linktocpage=true]{hyperref}
\usepackage[linktocpage=true]{hyperref}
\hypersetup{
colorlinks=true,
citecolor=ultramarine,
linkcolor=cadmiumgreen,
urlcolor=indigo(dye),
pdfauthor={},
pdftitle={},
pdfsubject={}
}

\def\lsim{\mathrel{\raise.3ex\hbox{$<$\kern-.75em\lower1ex\hbox{$\sim$}}}}
\def\gsim{\mathrel{\raise.3ex\hbox{$>$\kern-.75em\lower1ex\hbox{$\sim$}}}}

\def\mnras{Mon.\ Not.\ R.\ Astron.\ Soc.\ }
\definecolor{darkgreen}{cmyk}{0.85,0.2,1.00,0.2} 
\definecolor{purple}{cmyk}{0.5,1.0,0,0}

\begin{document}
	
\title{BAO Modulation as a Probe of Compensated Isocurvature Perturbations}

\author{Chen Heinrich}\email{chenhe@caltech.edu}
\affiliation{Jet Propulsion Laboratory, California Institute of Technology, Pasadena, California 91109}

\author{Marcel Schmittfull}
\affiliation{Institute for Advanced Study, Princeton, New Jersey 08540}

\begin{abstract}

Compensated isocurvature perturbations (CIPs) are opposite spatial fluctuations in the baryon and dark matter density. 
They can be generated for example in the curvaton model in the early Universe but are difficult to observe because their gravitational imprint nearly cancels. 
We therefore propose a new measurement method by searching for a spatial modulation of the baryon acoustic oscillation (BAO) scale that CIPs induce. 
We find that for a Euclid-like survey the sensitivity is marginally better than the WMAP cosmic microwave background (CMB) constraint, which exploits the CIP-induced modulation of the CMB sound horizon. 
For a cosmic-variance limited BAO survey using emission-line galaxies up to $z\sim7$ the sensitivity is between stage 3 and stage 4 CMB experiments.
These results include using CIP-galaxy cross-correlations, which improves the sensitivity by a factor of $\sim2-3$ for correlated CIPs. 
The method could be further improved with an optimal estimator, similarly to the CMB, and could provide a useful cross-check of other CIP probes.
Finally, if CIPs exist, they can bias cosmological measurements made assuming no CIPs.
In particular, they can act as a super-sample fluctuation of the baryon density and bias measurements of the BAO scale. 
For modern BAO surveys, the largest 2$\sigma$ CIP fluctuation allowed by Planck's 95\% bound could bias BAO measurements of $H(z)$ by 2.2\%, partially reducing the tension with the local $H_0$ measurements from 3.1$\sigma$ to 2.3$\sigma$.

\end{abstract}
\pacs{}
\maketitle

\section{Introduction}
\label{sec:intro}

Measurements of the cosmic microwave background (CMB) have shown that the Universe started with adiabatic initial conditions~\cite{Ade:2015lrj,Ade:2015ava,Akrami:2018odb} in which all particle species have the same fractional fluctuations in space. Isocurvature perturbations, on the other hand, are differences between the fractional fluctuations of two species and could arise from additional fields during inflation. Although the total matter-to-radiation isocurvature has been well constrained by \textit{Planck} to be less than a few percents~\cite{Akrami:2018odb}, there remains the possibility of an isocurvature mode orthogonal to that of the matter-radiation -- the compensated isocurvature perturbations (CIPs) -- which are still relatively unconstrained~\cite{Grin:2013uya, Munoz:2015fdv, Smith:2017ndr, Akrami:2018odb}.

In the CIP mode, the baryon and cold dark matter (CDM) have opposite fluctuations while the radiation has no perturbations~\cite{Grin:2011tf}. The result is that the total gravitational effects are canceled at linear order, leaving no matter-radiation isocurvature nor gravitational imprints (at least for scales above the baryonic Jeans scale). CIPs could arise from curvaton models in which an additional scalar field, the curvaton, is a spectator during inflation but later decays and seeds most of the curvature or adiabatic perturbations in the Universe~\cite{Linde:1984ti, Linde:1996gt, Langlois:2000ar,Lyth:2002my}. Depending on different decay scenarios of the curvaton, CDM or baryons can be produced by, before or after the curvaton decay, giving rise to different amounts of CIPs. 

The CIPs in the curvaton model are fully or partially correlated with the adiabatic perturbations depending on how much of the Universe's energy density is dominated by the curvaton at its decay. The largest possible fully correlated CIPs in the curvaton model would be about 16 times the adiabatic perturbations, corresponding to the curvaton scenario in which baryon number is produced by curvaton decay, and dark matter before the decay (see e.g.~\cite{He:2015msa}). 
 This leaves room for potential detection with a variety of cosmological probes.  

A few such probes of CIPs have been proposed so far, e.g.~gas fraction of galaxy clusters~\cite{Holder:2009gd}, primordial non-Gaussianities~\cite{Gordon:2002gv} and 21cm measurements~\cite{Gordon:2009wx} (see Ref.~\cite{Smith:2017ndr, Akrami:2018odb} for a more comprehensive review). One of the major methods to distinguish curvaton models is to search for the effects of a modulated sound horizon in the CMB. Even though the gravitational effects of the baryon fluctuations are compensated by those of dark matter in the CIP mode, there is still a spatial modulation of the baryon-to-photon ratio, leading to a modulation of the sound speed in the photon-baryon fluid before recombination. 
Such a modulation can be measured similarly to gravitational lensing of the CMB -- via a quadratic estimator or from the smoothing of acoustic peaks in the CMB power spectra.
Applying these methods to WMAP temperature data~\cite{Grin:2013uya} and \textit{Planck} CMB and CMB lensing power spectra~\cite{Akrami:2018odb, Smith:2017ndr, Valiviita:2017fbx} gave some of the most stringent constraints on CIP fluctuations to date. 

\begin{figure}[t]
\includegraphics[width=0.48\textwidth]{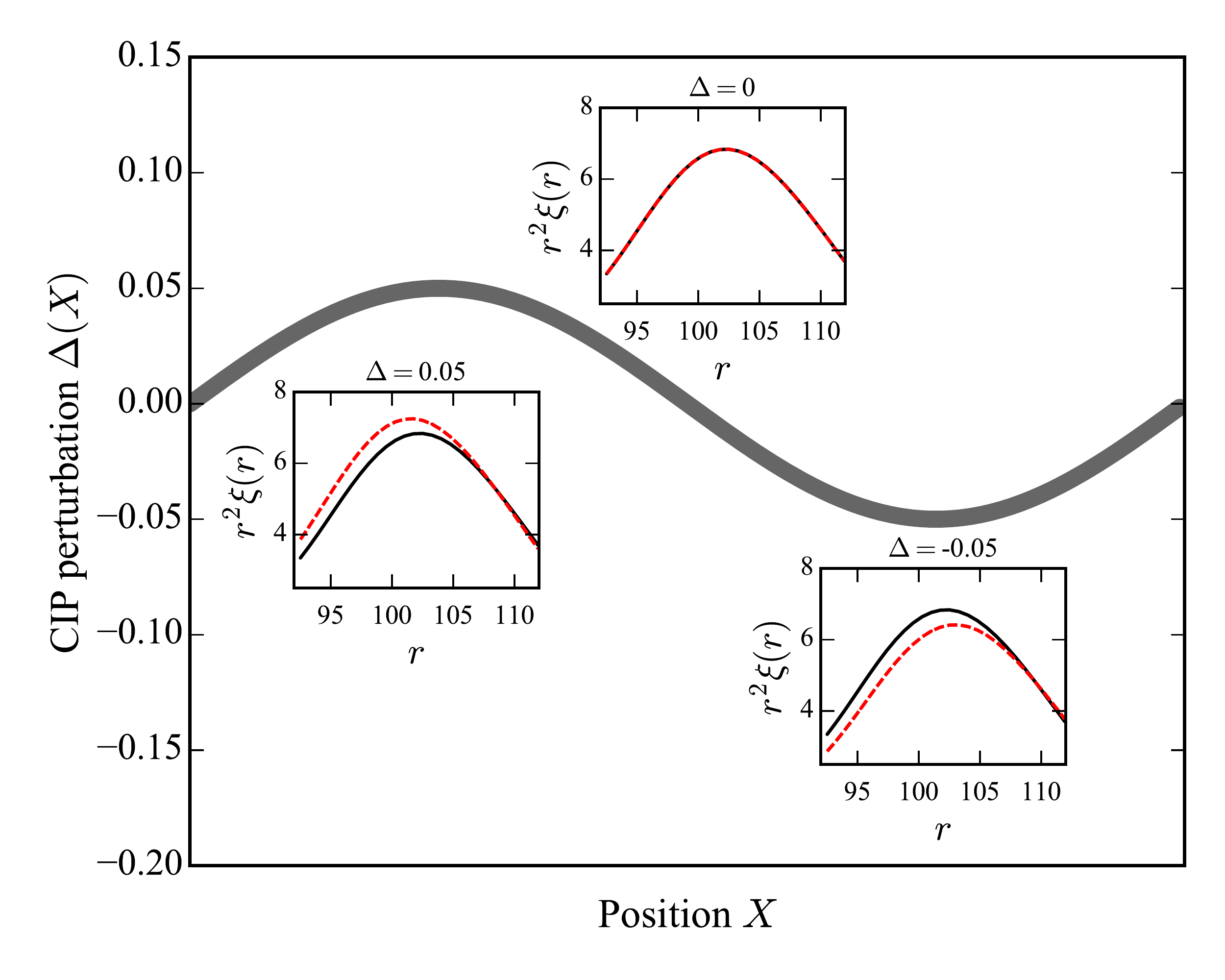}
\caption{1D illustration of the effect of a large scale CIP $\Delta(X)$ on the BAO peak in the local correlation function of galaxies $\xi(r)$ in the separate universe limit. The local baryon density in a subvolume is $\Omega_b=\bar{\Omega}_b(1+\Delta(X))$, and the local CDM density is compensated, $\Omega_c=\bar{\Omega}_c-\bar{\Omega}_b\Delta(X)$, leaving $\Omega_b+\Omega_c$ unchanged. 
The spatial modulation of the baryon-to-photon ratio changes the plasma sound speed before recombination and leads to a modulation of the BAO scale, observable in the local correlation function of galaxies (insets: dashed red with CIP, solid black without CIP). 
}
\label{fig:cartoon}
\end{figure}

We propose in this work a new method of looking for CIPs by noting that 
a modulated sound speed affects not only the acoustic peaks in the CMB, but also the clustering of galaxies: When baryons stop feeling the radiation pressure after recombination, the primordial sound wave stops, leading to an enhanced clustering of baryons (and galaxies) when separated by the distance the sound wave traveled until recombination --- the baryon acoustic oscillation (BAO) scale. 
A modulation of the sound speed therefore leads to a modulation of the BAO scale as a function of 3-dimensional position (see \fig{cartoon} for a simple illustration).
As a result, galaxy surveys can constrain CIPs by searching for such a modulation of the BAO scale.
We will outline for the first time a procedure for measuring CIPs from BAO surveys in this way, and forecast the errors on the amplitude of fully correlated CIPs for future surveys. 

We show that when the CIP-galaxy cross spectrum is included, a fiducial next-generation galaxy survey yields similar constraints to WMAP~\cite{Grin:2013uya}. We find that the cosmic-variance-limit (CVL) using $z\leq7$ emission line galaxies yields a constraint between stage 3 and stage 4 CMB experiments, which is about 3 times worse
 than the CMB CVL. We also study how the results depend on different properties of the surveys, such as survey volume, cell size, noise level, and inclusion of the cross-correlation with the galaxy density. Similarly to cross-correlating CIPs with CMB anisotropies as adopted in Ref.~\cite{He:2015msa}, we find that constraints for large future surveys are improved by a factor of about $\sim2-3$ when cross-correlations are included. 

The BAO method could cross-check CMB measurements, which are contaminated by gravitational lensing signals, and could potentially help to remove biases due to CMB lensing~\cite{Heinrich:2016gqe, Heinrich:2017psm}. We also note that since we are only interested in studying the feasibility of the BAO method, we will use a simple and non-optimal estimator for these forecasts that could in principle be improved with an optimal estimator.

We note that the authors of Refs.~\cite{Soumagnac:2016bjk,Soumagnac:2018atx} also proposed to use BAOs to probe CIPs.
They measure the luminosity-weighted galaxy density fluctuations and compare them to the unweighted ones, with the expectation that more stars are formed in regions of higher baryon content, indicative of CIPs. 
An advantage of this method is that it does not require dividing the survey into multiple patches.
But two modeling systematics are present in this approach: How galaxy luminosity and star formation depend on the baryon fraction of the halo, and how the baryons in the halos trace the total baryon content of the Universe. 
Our method is different and does not rely on modeling these nonlinear processes because it only uses the spatial modulation of the BAO scale, which is imprinted in the density fluctuations right after recombination and is transferred to the clustering of galaxies.
It would be interesting to combine both methods, especially for future large-volume surveys.

We begin in \refsec{background} by providing background on CIPs in the curvaton model and describing their effects on BAO measurements. In \refsec{recon}, we describe a reconstruction procedure for measuring CIPs from modulated BAOs. In \refsec{results} we use Fisher matrix to forecast the sensitivity of a fiducial BAO survey to correlated CIPs and study the forecast dependencies on different survey parameters. We discuss the results in \refsec{discussion} and conclude in \refsec{summary}. 

\section{CIPs and BAO modulation}\label{sec:background}

In this section we review what CIPs are and describe how they modulate the BAO scale and the observable galaxy power spectrum.

\subsection{CIPs}

The primordial perturbations in the early Universe can be decomposed into two types. The adiabatic fluctuations $\zeta$ are fluctuations on constant density slicings, while the isocurvature fluctuations $S_{i\gamma}$ are the differences between the fractional fluctuations of number density $n$ of species $i$ and the photons $\gamma$
	\begin{equation}
		S_{i\gamma}=\frac{\delta n_{i}}{n_{i}}-\frac{\delta n_{\gamma}}{n_{\gamma}}.
	\end{equation} 
Here $i \in \left\{b,c,\nu\right\}$ with $b$ for baryons, $c$ for cold dark matter and $\nu$ for neutrinos. 

The compensated isocurvature mode $\Delta$ is a special combination of those perturbations for which
the baryon number density fluctuations are exactly compensated by those of CDM
	\begin{eqnarray}
	S_{b\gamma} = \Delta ,\quad	S_{c\gamma} = -\frac{\rho_b}{\rho_c} \Delta, \quad
	S_{\nu\gamma} = 0,
	\end{eqnarray}
where $\rho_i$ is the energy density of species $i$.	

An early-universe model that generates perturbations of this form is the curvaton model. 
In this model, 
the curvaton is a spectator scalar field during inflation but later decays into other particles. When the Hubble scale becomes comparable to the mass of the curvaton,
the field starts to oscillate and its energy density redshifts like that of matter in the radiation era after inflation. As a result, the curvaton could dominate the energy density of the Universe at its decay and give rise to most of the adiabatic perturbations in the Universe. 

Depending on whether it decays into baryon number or CDM,
isocurvature perturbations are produced and are correlated with the adiabatic fluctuations. In particular, different amount of CIPs can be produced, so one can use them to distinguish the decay scenarios of the curvaton model (see e.g.~\cite{He:2015msa}).

In the case where the curvaton dominates the Universe's energy density completely at its decay, it would be responsible for generating all the adiabatic perturbations $\zeta$, and any resulting CIPs would be fully correlated with $\zeta$,
    \begin{equation}
\label{eq:AmplitudeDefn}
   \Delta = A \zeta,
    \end{equation}
where $A$ determines the amplitude of the CIP.
Our goal is to constrain this CIP amplitude $A$.

There are two decay scenarios that are particularly interesting because of their relatively large
CIP amplitude: $A \approx 3\Omega_c/\Omega_b \approx 16$ if baryon number is produced by the curvaton decay, and CDM before the decay, and $A= -3$ if CDM is produced by curvaton decay, and baryon number before the decay.

\subsection{BAO Modulation from CIPs} 
\label{se:ModulatedObs}

The CIP modes described above modulate the baryon-to-photon ratio spatially. As a result, the acoustic waves in the baryon-photon fluid generated by the adiabatic perturbations propagate with spatially varying sound speed before recombination. This modulation of the sound speed leads to a position-dependent BAO scale, which can be used to constrain the CIP amplitude as we are going to show.

How exactly do the CIPs modulate the BAO scale?
For CIP modes with wavelength shorter than the sound horizon at recombination, the sound waves speed up and slow down over many wavelengths of the CIP modulation, leading to a negligible net effect at recombination~\cite{He:2015msa}. But for CIP wavelengths longer than the sound horizon, 
sound waves propagate on a local background that is slowly modulated. In this limit, the behavior of subhorizon modes in the patch around a location $\VX$ can be approximated as if they were in a separate universe with perturbed cosmological parameters~\cite{Grin:2011tf}
	\begin{align} 
	\delta \Omega_b = \bar{\Omega}_b \Delta(\VX), \quad \delta \Omega_c = - \bar{\Omega}_b \Delta(\VX),
	\label{eq:separate}
	\end{align}	
where barred quantities denote values in the absence of CIPs.
From now on we will use $\VX, \VK$ to denote the long-wavelength CIP position and wavenumber, and use $\vx, \vk$ to denote those of the short-wavelength perturbations on the modulated background.

As a consequence, a small, long-wavelength CIP perturbation $\Delta(\VX) \ll 1$ changes the local mean baryon density as $\bar{\Omega}_b \rightarrow \bar{\Omega}_b (1+\Delta(\VX))$, and the sound speed of the baryon-photon fluid in the patch around $\VX$ changes accordingly as
\beq
\bar{c}_s \longrightarrow c_s(\VX) =\sqrt{\frac{1}{3}\left( 1 + \frac{3\bar{\rho}_b (1+ \Delta(\VX))}{4\bar{\rho}_{\gamma}}\right) ^{-1}} \approx \tilde{\alpha}(\VX)\,\bar c_s ,\\
\eeq
where we have Taylor expanded in $\Delta\ll 1$ to obtain
\bea
\tilde{\alpha}(\VX) = 1 + \frac{\Delta(\VX)}{C},
\label{eq:Delta_to_alpha}
\eea
where 
\beq
C \equiv -\frac{2(1+R)}{R}, \;\;\;R \equiv \frac{3\bar\rho_b}{4 \bar\rho_{\gamma}};
\eeq
in the fiducial cosmology described in \refsec{fiducial}, $C\approx-5.23$.
The corresponding change in the local BAO scale is 
\beq
\bar{r}_{\rm BAO}\;\longrightarrow\;  \tilde{\alpha}(\VX)\, \bar{r}_{\rm BAO}.  
\label{eq:rBAO}
\eeq 
Note that for the convenience of linearly relating $\tilde{\alpha}$ to $\Delta$, we have defined it differently than the common convention, $\alpha = 1/\tilde{\alpha}$, which would have given $\bar{r}_{\rm BAO}\longrightarrow ~\bar{r}_{\rm BAO}/\alpha(\VX)$ for Eq.~\ref{eq:rBAO} and $f_{\rm BAO}(k/\alpha(\VX))$ in Eq.~\ref{eq:Pmm} below.

\subsection{Observable Imprints in the Galaxy Power Spectrum}

Given the modulation of the BAO scale, the matter power spectrum in the patch around $\VX$ is 
\beq
P_{m}(k,z;\VX) = [1+f_{\rm BAO}(\tilde{\alpha}(\VX) k)]\, P^{\rm nw}_{m}(k,z),
\label{eq:Pmm}
\eeq
where we have split the matter power spectrum into a smooth nowiggle part without BAO wiggles, $P^{\rm nw}_{m}$, and an oscillatory wiggle part, $f_{\rm BAO}$, due to baryons \cite{Hamann:2010pw,Vlah:2015zda,Baumann:2017gkg}. 
Specifically, we use the wiggle power spectrum evaluated with CAMB \cite{Lewis:1999bs} for the fiducial cosmology, and the corresponding nowiggle power spectrum from the appendix of Ref.~\cite{Vlah:2015zda}. 
This nowiggle spectrum is constructed to have the correct low-$k$ and high-$k$ behavior, and the same density normalization $\sigma_8$ and velocity dispersion $\sigma_v$ as the power spectrum with BAO wiggles\footnote{This is achieved by constructing a family of spline curves for the nowiggle spectrum and forming a linear combination such that the integral constraints for $\sigma_8$ and $\sigma_v$ are satisfied \cite{Vlah:2015zda}.}.

In reality, we observe the galaxy rather than the matter power spectrum to measure the BAO scale from galaxy surveys, so additional effects from galaxy bias, redshift space distortion (RSD) and nonlinear evolution need to be taken into account. For extracting the BAO signal in a local patch around $\VX$, we model the galaxy power spectrum as
\bea
P_{gg}(k,z; \VX) = b_g^2(z) R(\mu, z) P^{\rm nl}_{m}(k,z, \VX), 
\label{eq:Pgg}
\eea
where
\beq
R(\mu, z) = (1+\beta(z) \mu^2)^2
\label{eq:RSD}
\eeq
is the linear RSD effect with $\mu = \mathrm{cos}({\boldsymbol{\hat{k}}} \cdot \mathbf{\hat{n}})$ being the angle between ${\boldsymbol{\hat{k}}}$ and the line of sight direction $\mathbf{\hat{n}}$. We use $\beta(z) = f(z)/b_g(z)$ with $f(z) \approx \Omega_m(z)^{0.55}$ and follow Ref.~\cite{2013ApJ...767...89M} to model the linear galaxy bias $b_g(z)$ 
for emission line galaxies as $b_g(z)\tilde{D}(z) = 0.84$, where $\tilde{D}(z=0) \equiv 1$ is the growth normalized at $z=0$. 

We model the impact of nonlinear evolution as a Gaussian damping of the oscillations in Fourier space with widths $\Sigma_{\bot}$ and $\Sigma_{\parallel}$ respectively for modes perpendicular and along the line of sight~\cite{2007ApJ...664..660E,Seo:2007ns}. The matter power spectrum including these non-linear effects is therefore
\beq
P_{m}^{\rm nl}(k,z; \VX) =   
\left[1+  f_{\rm BAO}(\tilde{\alpha}(\VX) k) \mathcal{D}(k,\mu, z) \right] P^{\rm nw}_{m}(k,z), 
\label{eq:Pmnl}
\eeq
where
\beq
\mathcal{D}(k,\mu,z) = e^{-\frac{1}{2}k^2\left[\Sigma_{\bot}^2 + \mu^2(\Sigma_{\parallel}^2 - \Sigma_{\bot}^2)\right]}.
\eeq
Following Ref.~\cite{Seo:2007ns}, we use $\Sigma_{\bot} = \Sigma_0 D(z)$ and $\Sigma_{\parallel} = \Sigma_{\bot} (1+f(z))$ where $\Sigma_0 = 12.4 \times (\sigma_8/0.9) \hMpc = 11.4 \hMpc$, and where $D(z)$ is the growth normalized to $1/(1+z)$ during matter domination. Note that a fraction $p_{\rm recon}$ of the non-linear damping effect can be removed by applying BAO reconstruction which moves observed galaxy positions using the large-scale flow derived from the galaxy density on large scales \cite{2007ApJ...664..675E}. To account for this possibility, we let $\Sigma_0 \rightarrow \Sigma_0 (1-p_{\rm recon})$ where the default $p_{\rm recon}$ = 0.5. 

CIP modes $\Delta$ therefore affect the galaxy power spectrum $P_{gg}$ by changing $\tilde{\alpha}$, which changes the frequency of oscillations of $f_{\rm BAO}$,
\beq
\frac{\partial P_{gg}}{\partial \tilde{\alpha}}  = b^2 R(\mu) \mathcal{D}(k,\mu, z) \frac{\partial f_{\rm BAO}(k)}{\partial \mathrm{ln}k}P^{\rm nw}_{m}(k,z), 
\eeq
so that
\beq
\frac{dP_{gg}}{d\Delta} = \frac{1}{C}\frac{\partial P_{gg}}{\partial \tilde{\alpha}}
\label{eq:dPgg_dD}
\eeq
in linear theory. 
Note that there are additional signatures of the CIP, corresponding to additional terms in Eq.~\ref{eq:dPgg_dD} of the form $\sum_{i} (\partial{P_{gg}}/\partial p_i) (\partial p_i/ \partial \Delta)$ where $p_i$'s are other parameters that depend on $\Delta$ including $p_i = \Delta$ itself. We choose to ignore those terms as well as derivative terms that arise from varying components in Eq.~\ref{eq:Pmnl} other than $f_{\rm BAO}$, as we expect Eq.~\ref{eq:dPgg_dD} to be the dominant and cleanest signal for now. 
For example, the broadband shape of the power spectrum at high $k$ also depends on the baryon fraction (e.g.~it is suppressed for high baryon density) and would be modulated by the CIP mode as a function of $\VX$. 
However, modeling the response of the high-$k$ broadband to CIP modes and accounting for observational systematics that could also modulate the broadband power seems challenging.
The difficulty of modeling the broadband can be seen from the fact that current BAO analyses compute the broadband using a polynomial template 
\cite{2012MNRAS.427.2132P,2012MNRAS.427.3435A,2017MNRAS.464.3409B}.
It could be that the modeling and understanding of systematics will improve in the future, and one could draw additional information from these other terms. But we will choose to ignore them for now, corresponding to a conservative forecast.

\subsection{BAO Modulation from Other Effects}
\label{sec:other_modulation}
Up to now we have assumed that CIPs are the only effect modulating the BAO scale. In reality, other effects like matter density modes can also modulate the BAO scale. 
Over- (under-) densities of very long modes $\delta_m^{L}(\VK)$ that are longer than the BAO scale, $K\lesssim 1/r_{\rm BAO}$, correspond to closed (open) separate universes. They contract (dilate) all scales inside that local patch by $\delta_m^{L}(\VK)/3$~\cite{2012PhRvD..85j3523S}; due to this dilation the BAO scale measured in these patches is also modulated\footnote{When measuring the global power spectrum, this dilation together with nonlinear growth leads to a small shift of the inferred BAO scale, because subvolumes with a shrinked BAO scale contribute more to the global power spectrum than subvolumes where the BAO scale is stretched \cite{2012PhRvD..85j3523S}.
}.

In contrast, displacements from modes smaller than the BAO scale, i.e.~modes with $k\gtrsim 1/r_{\rm BAO}$, perturb galaxies separated by the BAO scale by moving them towards or away from each other.
This again stretches or shrinks the BAO scale locally depending on the realization of those modes, but the effect averages out and only smears out the BAO peak in the correlation function.
Therefore, the very long modes (longer than BAO scale) can bias the CIP measurement, while the modes shorter than the BAO scale reduce the signal-to-noise ratio.

These effects can be avoided by applying 
 BAO reconstruction \cite{2007ApJ...664..675E} to the entire observed volume.
 The reconstruction
 removes the BAO modulation by moving galaxies by the Zel'dovich displacement computed from the observed  modes that modulate the BAO scale, which removes the bias and smearing of the BAO feature.
In practice, this process of BAO reconstruction is not perfect, for example due to noise in the estimated Zel'dovich displacement, so that a residual BAO modulation may be left.  This can lead to a residual bias on the CIP measurement which would have to be accounted for. We will ignore such a potential bias here and restrict ourselves to computing the statistical power with which the CIP amplitude can be constrained in absence of any bias.

An alternative procedure would be to apply BAO reconstruction locally in each subvolume and account for the dilation and growth from long modes by manually rescaling the density in each subvolume based on the observed long mode 
on scales larger than the subvolume. We describe this in Appendix~\ref{sec:bias_removal}. The two procedures are in principle equivalent in linear theory, but may differ in practice, depending on the reconstruction algorithm used. 

An additional modulation of the observed BAO peak in the correlation function is expected from gravitational lensing, which can magnify or shrink the physical BAO scale depending on the structures along the line of sight.
This leads to a damping of the BAO feature in the global correlation function at the few percent level at $z=2$ and less at lower redshift \cite{0702606}.
Compared to the modulation from bulk flows and the induced damping of the global BAO feature, this BAO lensing modulation is therefore subdominant, and we will ignore it for simplicity.\footnote{If the BAO lensing modulation turns out to be sufficiently large to be worrying for CIP constraints, one could delens the BAO modulation, e.g., using external delensing based on observations of low-redshift galaxies, or using internal delensing that removes the lensing-induced non-Gaussianity of the observed BAO modulation.}

In sum, the CIPs change the sound horizon as a function of position on the sky and induce a large-scale modulation of the measured BAO scale. 
The contaminating modulation due to long wavelength modes can be avoided by applying BAO reconstruction, and the modulation due to gravitational lensing is small so that it can be ignored.

\section{Measuring CIPs using BAO}
\label{sec:recon}

The modulation of the BAO by CIP modes described above suggests the following procedure for the measurement of CIPs from the modulation of the BAO on the sky.

\begin{enumerate}
\item{Apply BAO reconstruction in the total survey volume.}

\item{Divide the total survey volume $V_{\rm tot}$ into $N_{\rm cell}$ cells with the same volume $V_{\rm cell}$, centered at positions $\VX_i$, where $i = 1, \dots, N_{\rm cell}$.}

\item{Measure the BAO scale $\hat{\tilde{\alpha}}(\VX_i)$ using galaxies in each cell $\VX_i$, where $\tilde{\alpha}$ is defined as the measured BAO scale $\hat{r}_{\rm BAO}$ with respect to the unmodulated BAO scale $\bar{r}_{\rm BAO}$, approximated by that of the fiducial cosmology measured by \textit{Planck}.} 

\item{Estimate the CIP density perturbation $\hat\Delta(\VX_i)$ from $\hat{\tilde{\alpha}}(\VX_i)$.
}

\item{Compute the CIP auto-spectrum $\hat{P}_{\Delta\Delta}(K)$ and its cross-spectrum $\hat{P}_{g\Delta}(K)$ with the long-wavelength galaxy density $\delta_{g}^{L}(\VK)$. Use them to measure the CIP amplitude $A$ in the curvaton model.} 
\end{enumerate}

In step 1, we apply BAO reconstruction to the entire survey to avoid contamination from BAO scale modulations induced by dilation and bulk flows, as discussed in \refsec{other_modulation}.

In step 2, the survey volume needs to be divided into sufficiently many $N_{\rm cell}$ cells for the Fourier transform of the later steps to make sense (i.e., to see a modulation). At the same time, we cannot use arbitrarily many cells
 because each cell must be larger than the BAO scale so that we can detect a BAO signal in each cell. We will explore the effect of cell size in Sec.~\ref{sec:dependencies}.

Once the BAO map $\hat{\tilde{\alpha}}(\VX_i)$ is obtained in step 3, the CIP estimator is computed in step 4 as
\beq
\hat{\Delta}(\VX_i) = C \left(\hat{\tilde{\alpha}}(\VX_i) - 1\right),
\eeq
where the unmodulated BAO scale $\bar r_{\rm BAO}$ can be taken as that measured by the CMB. 
Taking the power spectrum in step 5 implicitly involves taking the Fourier transform of $\hat{\Delta}(\VX_i)$. Since the true CIP map is multiplied by the survey window function in the measurement, the measured $\hat{\Delta}(\VK)$ is the truth $\Delta(\VK)$ convolved with the Fourier transform of the window function plus noise. 

The CIP measurement noise is directly related to the BAO error per cell. Assuming uncorrelated and homogeneous noise, we can characterize the noise by its power spectrum alone
\beq
\label{eq:NDD}
N_{\Delta\Delta}(K) = V_{\rm cell} \,\sigma_{\Delta}^2 \, \left[W(KR)\right]^{-2}.
\eeq
The low-$K$ amplitude of the noise is proportional to the cell volume $V_{\rm cell}$ multiplied by $\sigma_{\Delta}^2 = C^2 \sigma_{\tilde{\alpha}}^2$, where $\sigma_{\tilde{\alpha}}$ is the BAO measurement error in a cell.
The product $V_{\rm cell} \,\sigma_{\Delta}^2$ is approximately independent of cell size as long as there are enough cells and they are larger than the BAO scale.

The noise increases rapidly
at high $K$ because we cannot probe fluctuations smaller than the cell size, so in principle one would low-pass filter the $\hat{\Delta}(\VX_i)$ map before applying the Fourier transform in order to avoid aliasing. 
We model the low-pass filter as a convolution of the true map with a tophat window function with radius $R = (4\pi/3)^{1/3} r_{\rm cell}$ such that the volume of the sphere is the same as the cell volume, where
\beq
r_{\rm{cell}} = V_{\rm cell}^{1/3}.
\eeq
This filtering corresponds to multiplying the signal in Fourier space by 
\beq
\mathrm{W}(KR) = \frac{3}{(KR)^3}[\,\mathrm{sin}(KR) - KR\,\mathrm{cos}(KR)\,].
\eeq
We checked that our results do not change significantly if we use a Gaussian convolution instead, where $N_{\Delta\Delta}(K) = V_{\rm cell} \,\sigma_{\Delta}^2 e^{(K /(K_{\rm Nyquist}))^2}$ such that frequencies above the Nyquist frequency $K_{\rm Nyquist} = \pi/r_{\rm cell}$ are suppressed.

The procedure outlined above is effectively measuring a position-dependent galaxy correlation-function where the BAO peak shifts depending on the position on the sky. More optimally, a quadratic estimator $\hat{\Delta}^{\rm mv}(K)$ can be constructed using off-diagonal correlations of the galaxy densities with weights that minimize the estimator variance. By definition the minimum variance estimator would have better noise properties than the one we use in this paper. 
Computing such an optimal estimator is however beyond the scope of this work. We will leave it for future work and focus on first assessing the method using the simpler estimator described above.

Since we consider curvaton CIPs, which are fully correlated with the adiabatic fluctuations $\zeta$, we can measure cross-correlations of $\Delta$ with other probes of $\zeta$ in step 5 to improve the sensitivity in the low signal-to-noise regime. 
Specifically, we consider the cross-correlation of $\Delta$ with the galaxy density which traces $\zeta$.

For the CMB, the authors of Ref.~\cite{He:2015msa} found that including cross-correlations with CMB temperature and polarization anisotropies leads to an improvement by a factor of $\sim 2-3$ in $\sigma_A$ for CIP measurements using quadratic estimators of the CMB. We shall see in Sec.~\ref{sec:results} that a similar improvement applies when we include cross-correlations with galaxies in the BAO modulation method.

\section{Fisher Forecast}\label{sec:results}

In this section, we describe how we forecast
the sensitivity of the BAO method to the amplitude of curvaton CIP modes
assuming a fiducial survey (\refsec{fiducial}), and explore how the results depend on various aspects of the data and assumptions (\refsec{dependencies}).

\subsection{Fiducial Survey}\label{sec:fiducial}

We will use as observables the auto- and cross-spectra of the measured CIP mode $\Delta$ and the galaxy density on large scales
\begin{align}
	P_{XY}(K,z) = T_{X}(K,z) T_{Y}(K,z)
	P_{\zeta\zeta}(K),
\label{eq:PXY}
\end{align}
where 
\beq
P_{\zeta\zeta}(K) =  2\pi^2 K^{-3} A_s (K/K_0)^{n_s-1}, 
\eeq
with $X, Y \in \{g,\Delta\}$ where the CIP transfer function is
\begin{align}
T_{\Delta}(K) = A,
\end{align}
and the galaxy transfer function on scales larger than the BAO scale is
\begin{align}
T_{g}^{L}(K,z) = b_g(z) T_m^{\rm nw}(K,z).
\end{align}
Note that we ignore RSD effects and use $T_m^{\rm nw} \equiv \sqrt{P_{m}^{\rm nw}/P_{\zeta\zeta}}$ from the no-wiggle matter power spectrum
because we do not expect BAO wiggles on these scales.
Recall that we only need to cross-correlate galaxy density modes on the same scales as the CIPs, and these are larger than the BAO scale by construction (one can only measure the modulation of the BAO scale on scales larger than the BAO scale). 

We use the Fisher information matrix $F_{ij}$ to predict the inverse covariance matrix of a set of parameters $p_i$ that determine the reconstructed CIP field and the galaxy number density field. Under the assumption of Gaussian statistics for these fields, it can be approximated as 
\bea
F_{ij} &=&\int_{K_{\rm min}}^{K_{\rm max}} dK \frac{4\pi K^2 V_{\rm survey}}{(2\pi)^3}  \int_{-1}^{1} \frac{d\mu}{2}  \notag\\
& \times& \sum_{a,b} \frac{\partial \tilde{P}_{a}(K,z_c)}{\partial p_i}  \left[\mathcal{C}^{-1}(K,z_c) \right]_{ab}\frac{\partial \tilde{P}_{b}(K,z_c)}{\partial p_j} 
, 
\label{eq:Fisher}
\eea
where we assume a central redshift $z_c$ at which the power spectrum, noise and covariance matrix are calculated (we will suppress the $z_c$ dependence from now on). The covariance matrix is \beq
\mathcal{C}_{XX', YY'} = \tilde{P}_{XY} \tilde{P}_{X'Y'} \,+\, \tilde{P}_{XY'}\tilde{P}_{X'Y},
\eeq
where 
\beq 
\tilde{P}_{a}(K) = P_{a}(K) * W(K)+ N_{a}(K)
\eeq
are the observed power spectra for $a \in \{\Delta\Delta,\, g\Delta,\, gg\}$ where $P_{a}(K)$ are defined in Eq.~\ref{eq:PXY}, $N_{\Delta\Delta}$ is defined in \eqq{NDD}, $N_{gg} = 1/\bar{n}$ is the usual Poisson noise, we assume $N_{g\Delta}=0$ on these large scales,
and $W(K)$ is the survey window function in Fourier space. We set $\beta = 0$ in the modeling of $P_{gg}$ for the CIP Fisher matrix, but will include RSD when estimating the error of the BAO scale per cell. 

The effect of having a finite-sized survey is that the neighboring $K$ modes in the power spectrum are convolved by a window function with the width roughly equal to $dK \approx K_{\rm min}$. 
We therefore approximate the Fisher matrix above as a discrete sum with bin-size $dK = K_{\rm min}$ so that different $K$-bins are roughly uncorrelated and the covariance matrix is diagonal in $K$-space. For the wavenumber range we choose $K_{\rm min} = 2\pi V_{\rm survey}^{-1/3}$ which is limited by the survey volume, and $K_{\rm max} = \pi V_{\rm cell}^{-1/3}$, the Nyquist frequency set by the sampling frequency due to the finite cell size. 

We are interested in the CIP amplitude parameter $p_i = A$ defined in Eq.~\ref{eq:AmplitudeDefn}. The information on $A$ comes from the auto-spectrum of the reconstructed CIP, $P_{\Delta\Delta} \propto A^2$, and its cross-spectrum with the galaxy density $P_{g\Delta}\propto A$. Since the other cosmological parameters are well determined, and we do not expect them to be degenerate with $A$ as they do not modulate the BAO scale, we will fix them to the fiducial LCDM values that we assume. The error on $A$ is then
\beq
\sigma_{A} = \frac{1}{\sqrt{F_{A A}}}. \\
\eeq
We compute this error as a function of the fiducial value of $A$ and report the result when $A/\sigma_A=2$, corresponding to the $2\sigma$ detection threshold.

The BAO error $\sigma_{\tilde{\alpha}}$ per cell that goes into the calculation of $N_{\Delta\Delta}$ is obtained with a similar Fisher matrix, but with $a = gg$ and $p_i = \tilde{\alpha}$ (the BAO scale relative to the fiducial value).
In addition, the observed galaxy power spectrum includes redshift-space distortion effects as in Eqs.~\ref{eq:Pgg} and \ref{eq:RSD}.
For the wavenumber range, we use 
$k_{\rm min} = 2\pi V_{\rm cell}^{-1/3}$ and $k_{\rm max}= 0.6\,\ihMpc$. 
To use a diagonal covariance matrix, we again approximate the integral with a discrete sum of width $dk = k_{\rm min} = 2\pi V_{\rm cell}^{-1/3}$. Since $\partial \tilde{P}_{gg}(k)/\partial \tilde{\alpha}$ has wiggles with period roughly $1/r_{\rm BAO}$, it is no longer slowly varying inside each bin for small cell size close to the BAO scale $r_{\rm cell} \sim r_{\rm BAO}$. 
To deal with this, we do not use bin centers for the summand but approximate the observed power spectrum derivative by averaging
\beq
\frac{\partial P_{gg}(k, \mu)}{\partial \tilde{\alpha}} D(k, \mu)
\eeq
within a $k$-bin. Note that the power spectrum itself is still slowly varying inside each bin, so it is safe to approximate the covariance matrix with its value at the bin-center. We have checked that using a no-wiggle power spectrum for the covariance computation does not change our results.

For surveys with large redshift range, the sensitivity of the BAO measurement would also vary with redshift, and the approximation of homogeneous CIP reconstruction noise would break down. In addition, the power spectrum and covariances would also evolve with redshift. However, we will take a fiducial survey with a sufficiently small redshift range and ignore this effect in the fiducial forecast.
For simplicity we will also assume this when we extrapolate results to larger volumes in Sec.~\ref{sec:dependencies}. 

Throughout, we use a flat $\Lambda$CDM cosmology consistent with \textit{Planck} 2015 results~\cite{Ade:2015xua}, with $\Omega_b h^2= 0.0223$ and $\Omega_c h^2 = 0.1188$, the adiabatic scalar power spectrum with amplitude $A_s = 2.207\times10^{-9}$ at pivot wavenumber $k_0 = 0.05\, \mathrm{Mpc^{-1}}$, spectral index $n_s = 0.9645$, Hubble constant $h = 0.6711$, resulting in $\sigma_8 = 0.8249$ assuming one massive neutrino with $m_{\nu}~=~0.06 \,\mathrm{eV}$.

We run the forecast described above for a fiducial galaxy survey with sky fraction $f_{\rm sky} = 0.5$, signal-to-noise ratio $\bar{n}P_{0.2} = 5$ where $P_{0.2}$ is evaluated at $k=0.2~\ihMpc$ and $\mu =0$, reconstruction fraction $p_{\rm recon} = 0.5$ and a single redshift bin from $z = 0.75$ to $1.25$ with bin center $z_c = 1.0$, corresponding to a survey volume of $V_{\rm survey} = 27.8\, (\hGpc)^3$ which we divide into $N_{\rm cell} = 1000$ cells. 
The number density $\bar{n}$ of this fiducial survey is similar to what would be expected from next-generation galaxy surveys with flux limit of $1-2 \times 10^{-16} \mathrm{erg\, s^{-1} cm^{-2}}$~\cite{Merson:2017efv}, similar to Euclid~\cite{EuclidWhitePaper} at $z \sim 1$, but with a smaller redshift range and larger $f_{\rm sky}$. 
We find that the BAO method applied to this fiducial survey can detect the CIP amplitude with $2\sigma$ if $A=484$
(see Table~\ref{tab:2sigma}). This is similar to WMAP's sensitivity to CIPs without cross-correlations~\cite{Grin:2013uya, He:2015msa} (see Appendix~\ref{sec:conversion} for the conversion of WMAP results and the subtleties involved).

\begin{table}[htbp]	
\caption{$2\sigma_A$ detection threshold for the correlated CIP amplitude $A$ given the auto-spectrum of the CIP modes reconstructed from BAO modulation, the auto-spectrum of the galaxy density on large scales and their cross-spectrum,
for the fiducial next-generation survey and a cosmic-variance-limited survey of emission-line galaxies (see \refsec{fiducial} for assumptions). 
The fiducial survey performs similarly to current WMAP~\cite{Grin:2013uya} (see Appendix~\ref{sec:conversion}), whereas the CVL result falls between a stage-3 ($2\sigma_A \sim 60$) and stage-4 ($2\sigma_A \sim 15$) CMB experiment if the CMB lensing bias effect is included.}
\label{tab:2sigma} 
\setlength{\tabcolsep}{1.5em}
\begin{ruledtabular}
\begin{center}
\begin{tabular}{ll}
          Survey & $2\sigma_A$ \\ \hline     	\noalign{\smallskip}
Fiducial  & 484         \\  
CVL (z = 7)      & 30     \\  
\end{tabular}
\end{center}
\end{ruledtabular}
\end{table}	

\subsection{Forecast Dependencies} \label{sec:dependencies}

How can the fiducial BAO survey be improved to increase the sensitivity to CIPs?  To examine this 
we vary various aspects of the data and forecast assumptions: The galaxy number density $\bar{n}$, the cell volume $V_{\rm cell}$ and the survey volume $V_{\rm survey}$. Unless specified otherwise, we vary the assumptions one at a time from the fiducial choices described in the last section. \\

\textit{Galaxy number density} -- 
We begin by varying the galaxy number density. For a given reconstruction fraction, the CIP detection threshold is close to cosmic variance limited when the galaxy power signal-to-noise ratio exceeds $\bar{n}P \approx$ a few, so there is little gain with improving $\bar{n}P$ in that regime. But for $\bar{n}P$ smaller than a few, improving $\bar{n}P$ makes a bigger difference. This is especially true when cross-correlations are included as improving $\bar{n}P$ decreases the variance of the observed $P_{g\Delta}$ as well. \\

\begin{figure}
\includegraphics[width=0.9\linewidth]
{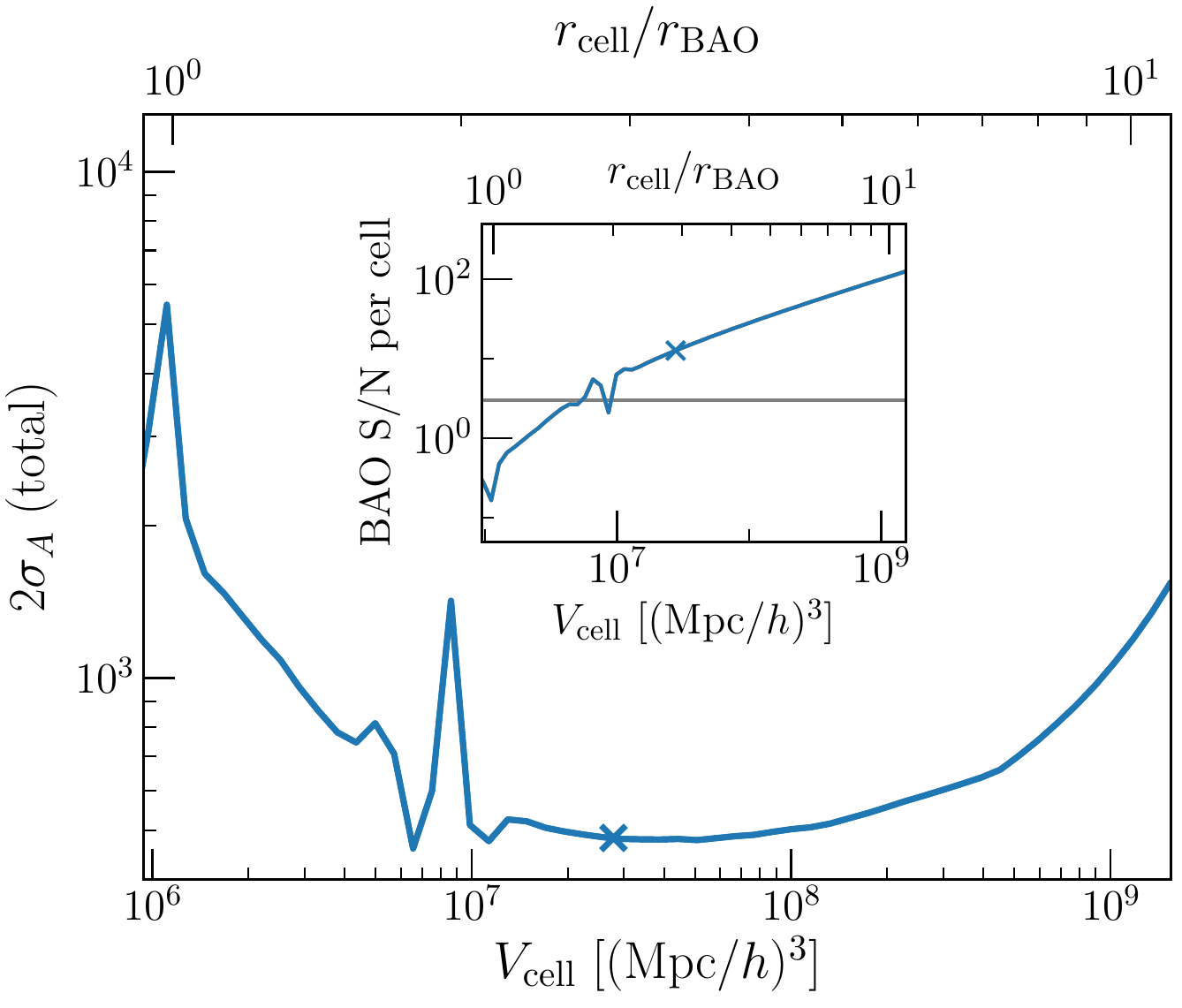}

\caption{$2\sigma$ CIP detection threshold in the main panel and BAO S/N per cell $\tilde{\alpha}/\sigma_{\tilde{\alpha}}$ in the inset as a function of cell volume (bottom axis) and the cell side length $r_{\rm cell}$ in units of $r_{\rm BAO}$ (top axis). We vary $K_{\rm max}$ = $\pi V^{-1/3}_{\rm cell}$ accordingly and fix all other parameters at their fiducial values (e.g. $V_{\rm survey} = 27.8\, (\hGpc)^3$). For these choices, there is a small optimal range of cell volume around side length $r_{\rm cell} \sim 3-4\, r_{\rm BAO}$. Note that the fiducial cell volume $V_{\rm cell, fid}$ given $N_{\rm cell} = 1000$ (blue cross) is already close to optimal. The spikes at 1 and 2 $r_{\rm BAO}$ are not physical but artifacts of an approximation made in the forecast (see text). }
\label{fig:cell_size}
\end{figure}

\textit{Cell volume} -- 
Next we study the dependence on the volume of the cells into which we divide the survey.
We vary 
$K_{\rm max} = \pi V^{-1/3}_{\rm cell}$ 
accordingly but fix all other parameters to their fiducial values. 
In Fig.~\ref{fig:cell_size} we plot the $2\sigma$ detection threshold vs cell volume on the bottom axis, and on the top axis, the cell side length $r_{\rm cell}$ in units of $r_{\rm BAO}$. If the cells are too small, there are not enough galaxy pairs separated by the BAO scale in each cell to make a high S/N measurement of the BAO scale: The inset shows that the BAO S/N per cell $\tilde{\alpha}/\sigma_{\tilde{\alpha}}$ drops below 3 for $r_{\rm cell} \lesssim 3\, r_{\rm BAO}$. On the other hand, if the cells are too big and approach the survey volume, only a small range of CIP wavenumbers is accessible, and the CIP detectability also degrades. 
This means that there is an optimal regime in the middle that is best for CIP detectability. For the fiducial choices used, the optimal cell size is around $r_{\rm cell} \sim 3-4\, r_{\rm BAO}$, and our fiducial choice of $V_{\rm cell} = 0.0278\, (\hGpc)^3 \approx (3\, r_{\rm BAO})^3$ given $N_{\rm cell} = 1000$ is nearly optimal. But this would change depending on survey properties. In general, with larger surveys comes more freedom to choose bigger cells instead of smaller but more cells without sacrificing CIP detectability; but for smaller surveys, such flexibility would be limited.

Note that the spikes in Fig.~\ref{fig:cell_size} are not physical, but are artifacts of the approximations used for obtaining uncorrelated $k$-bins of the galaxy power spectrum from which the BAO is measured. 
These spikes occur for $r_{\rm cell} = r_{\rm BAO}$ and $2\,r_{\rm BAO}$, where $k$-bins are spaced by $2\pi/r_{\rm BAO}$ or $\pi/r_{\rm BAO}$ and exactly one or half an oscillation of the signal $\partial P(k)/\partial \tilde{\alpha}$ averages out.
 We see this more clearly in the inset where the BAO S/N per cell $\tilde{\alpha}/\sigma_{\tilde{\alpha}}$ drops sharply for the same cell sizes. This is only an artificial consequence of using bins with $dk = k_{\rm min}=2\pi (V_{\rm cell})^{-1/3}$ and the averaged power spectrum derivative for each bin. In reality, the BAO detectability is a smooth function of the cell size as the power spectrum is convolved with the cell window function. \\

\begin{figure}
\includegraphics[width=0.9\linewidth]
{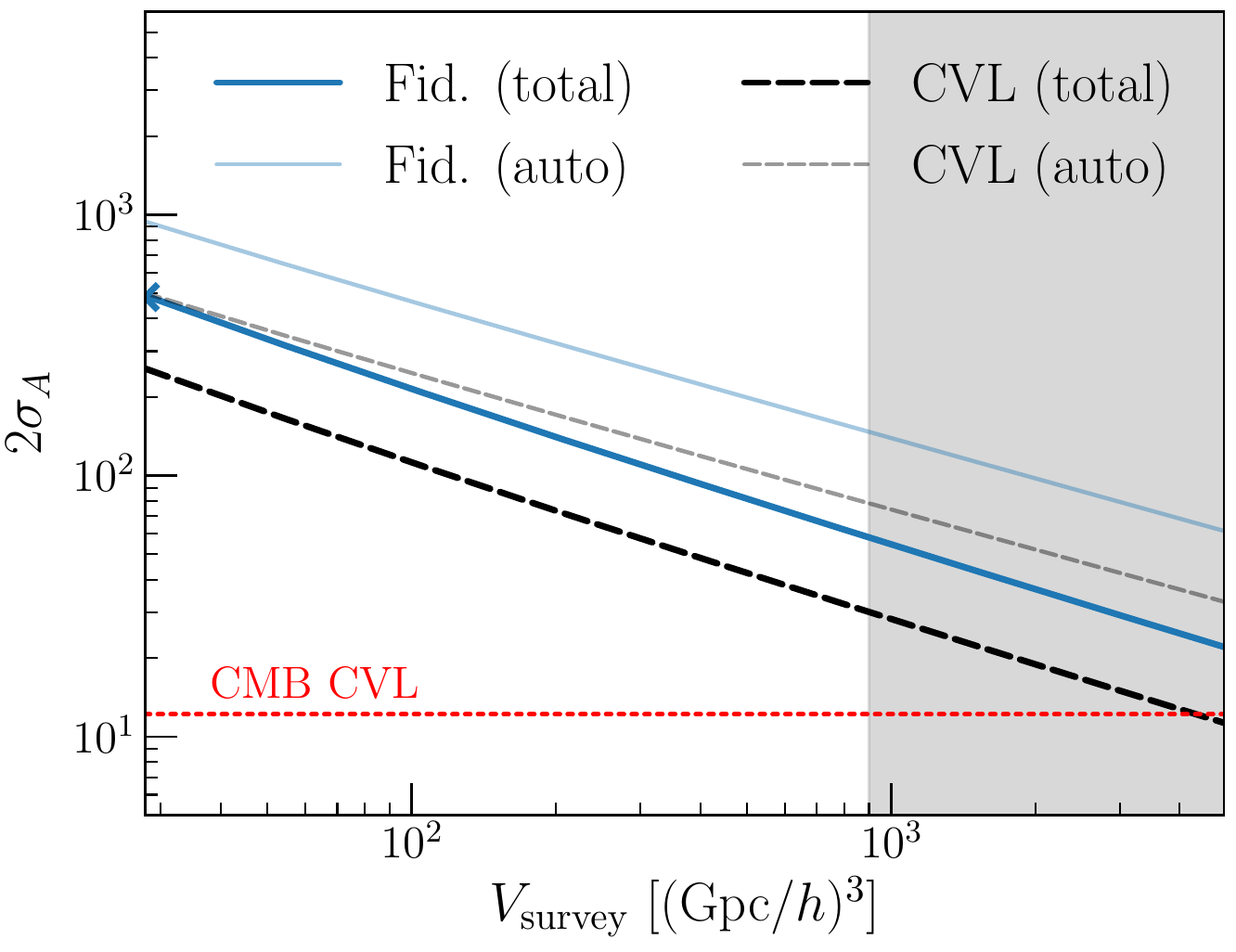} 
\caption{$2\sigma$ CIP detection threshold as a function of survey volume $V_{\rm survey}$ for fixed $V_{\rm cell} =0.0278\, (\hGpc)^3$. CIP auto-correlation only results are shown (thin lines) as well as the total results including the cross-correlation with galaxies (thick lines), for both the CVL (black dashed) and the fiducial survey (blue solid). 
Note that the CVL survey has a larger cell volume with side $r_{\rm cell} = 4 r_{\rm BAO}$.
For reference, we also show the CMB CVL for $l_{\rm max} = 2500$ (red dotted) which includes the degradation from lensing bias~\cite{Heinrich:2017psm}. Note that the CMB result is for an optimal estimator while the BAO is not. Compared to the CMB, the BAO CVL experiment with $z_{\rm max} \sim 7$ is about 3 times worse. For reference, the volumes for $z_{\rm max} = 2, 7, 150$ are $\{0.19,\, 0.9,\, 3.0\}\times 10^{3}\,(\hGpc)^3$ respectively for $f_{\rm sky} = 1$, and the right edge is at $z_{\rm max} = 1100$ in the fiducial cosmology.  } 
\label{fig:vsurvey}
\end{figure}

\textit{Survey volume} -- 
Next we examine the dependence of $2\sigma_{A}$ on the survey volume. We vary correspondingly $dK=K_{\rm min} = 2\pi (V_{\rm survey})^{-1/3}$, but keep the cell volume and $K_{\rm max}$ fixed.
The result is shown in Fig.~\ref{fig:vsurvey}. In addition, for the CVL survey, instead of using the fiducial cell volume, we fix it at a slightly more optimal choice of $V_{\rm cell} = (4\,r_{\rm BAO})^3$. Note also that although the volume is scaled up, all quantities in the Fisher matrix are still being calculated at fixed redshift $z_c=1$. A few subtleties are involved with this assumption. For a realistic survey with fixed flux limit, one expects the galaxy number density to decrease with redshift, increasing the Poisson noise. On the other hand, the galaxy bias also increases with redshift and enhances the signal. Finally, the percentage of BAO reconstruction achievable grows closer to 100\% at higher redshift since structure formation is less nonlinear there. 

None of these effects matter for the CVL survey, which has zero Poisson noise and 100\% reconstruction by definition. Furthermore, for cross-correlation dominated results the time evolution of the galaxy bias in the signal and the covariance pieces would cancel each other in absence of Poisson noise. Therefore we do not take into account any redshift evolution when we compute the CVL results, which is our main interest here. But we do show the same calculation for the fiducial survey as well, which is not meant to be accurate but only provides a rough comparison with CVL.

In Fig.~\ref{fig:vsurvey}, we show the $2\sigma$ detection threshold for the CVL (black dashed) and fiducial (blue solid) experiments using the auto-correlation only (thin lines) 
and including cross-correlations (thick lines). With the cross-correlation included, the fiducial experiment performs nearly as well as the CVL experiment with the auto-correlation alone. Moreover, we find that most of the improvement from the fiducial to the CVL curve at the same volume comes from improving the BAO reconstruction fraction from 0.5 to 1. So for a survey with fixed volume and $\bar{n}P$, its ability to detect CIPs can still improve with improved BAO reconstruction algorithms (e.g.~\cite{2017PhRvD..96l3502Z,2017ApJ...841L..29W,2017PhRvD..96b3505S,2017JCAP...12..009S,2018PhRvD..97b3505S,2018MNRAS.478.1866H,2019MNRAS.484.3818S}).

While the sensitivity to CIPs keeps improving with larger volume,  there are actually no galaxies during the dark ages, so one cannot in reality use all the volume up to the surface of last scattering. For $z_{\rm max} \sim 7$, corresponding to $V_{\rm survey} \sim 900\, (\hGpc)^3$, the BAO CVL with a slightly more optimal cell size $r_{\rm cell} = 4\, r_{\rm BAO}$ yields $2\sigma_A = 30$. This sensitivity is between stage-3 and stage-4 CMB experiments which would have $2\sigma_A \sim 60$ and $2\sigma_A \sim 15$ respectively assuming they suffer from a similar (if not worse) factor of 1.5 degradation than the CVL experiment due to lensing bias effects. Compared to the CVL CMB experiment, the galaxy BAO CVL at $z \lesssim 7$ is a factor of 2.5 (3.6) worse than the CVL CMB experiment with (without) degradation from lensing bias. 

To understand the relative strength of the CMB and BAO methods in their current forms it is useful to make an order of magnitude estimate.
As an indication of CIP detectability we consider the S/N of the acoustic scale per patch and the number of patches on the last scattering surface and in the volume up to $z_{\rm max} = 7$. For the CMB, the fractional uncertainty of the angular size of the sound horizon $\theta_{*}$ is $\Delta \theta_*/\theta_* = 4.4\times10^{-4}$ in the \textit{Planck} 2015 measurement, which would be improved by a factor of~7 by a full-sky CVL experiment~\cite{Scott:2016fad}. 
Using patches large enough to include the first acoustic peak, say square patches with side length $2.7\, r_{\rm BAO}$, 
the S/N per patch would be~$\sim133$
 using the naive $1/\sqrt{N}$ scaling with $N = 1.4 \times 10^4$.
Now a similar number of cubic cells with side length $4\, r_{\rm BAO}$
 exists in the volume between $z = 0$ and 7. For emission line galaxies, the CVL BAO S/N would be about $\approx$ 44
 per patch, which is a factor of~$\sim$3 lower than the CMB, so we do expect slightly better performance from the CMB. We keep in mind that this is only an order of magnitude estimate, as the CIP detectability also depends on other factors such as the optimization of the estimator, cross-correlation coefficient between the spectra, etc.

Despite this expected advantage of the CMB method, the BAO method that we assumed for the CIP forecasts above can be improved in a number of ways.
First, similarly to the CMB forecast in Fig.~\ref{fig:vsurvey}, the BAO method could be modified to use an optimal quadratic estimator.
This would involve deriving weights that minimize the variance of a quadratic estimator built from the correlation between different Fourier modes  of the galaxy density.
Another possible improvement unique to the BAO side is to use multiple types of galaxies as biased tracers of the adiabatic perturbations to cancel some sample variance; there are also other BAO probes to take advantage of such as those from Lyman-$\alpha$ measurements.
Furthermore, 21cm experiments have the potential to offer competitive BAO measurements in the future, and could extend the range of galaxy surveys to higher redshift~\cite{Ansari:2018ury}. The 21cm signal is there, in principle, all the way until $z \sim 150$, though foregrounds could be more challenging at higher redshifts. The 21cm observations could serve as additional BAO measurements, or as a probe of the adiabatic perturbations. 
Finally, the forecast could be improved using CIP information from the modulation of the broadband power spectrum. 

\subsection{Impact of Including Cross-Correlation}

In this section we examine in more detail how the error on the CIP amplitude $\sigma_A$ benefits from including the cross-correlation between the reconstructed $\Delta$ field and the galaxy density, for different values of the CIP amplitude $A$.
\begin{figure}
\includegraphics[width=0.9\linewidth]{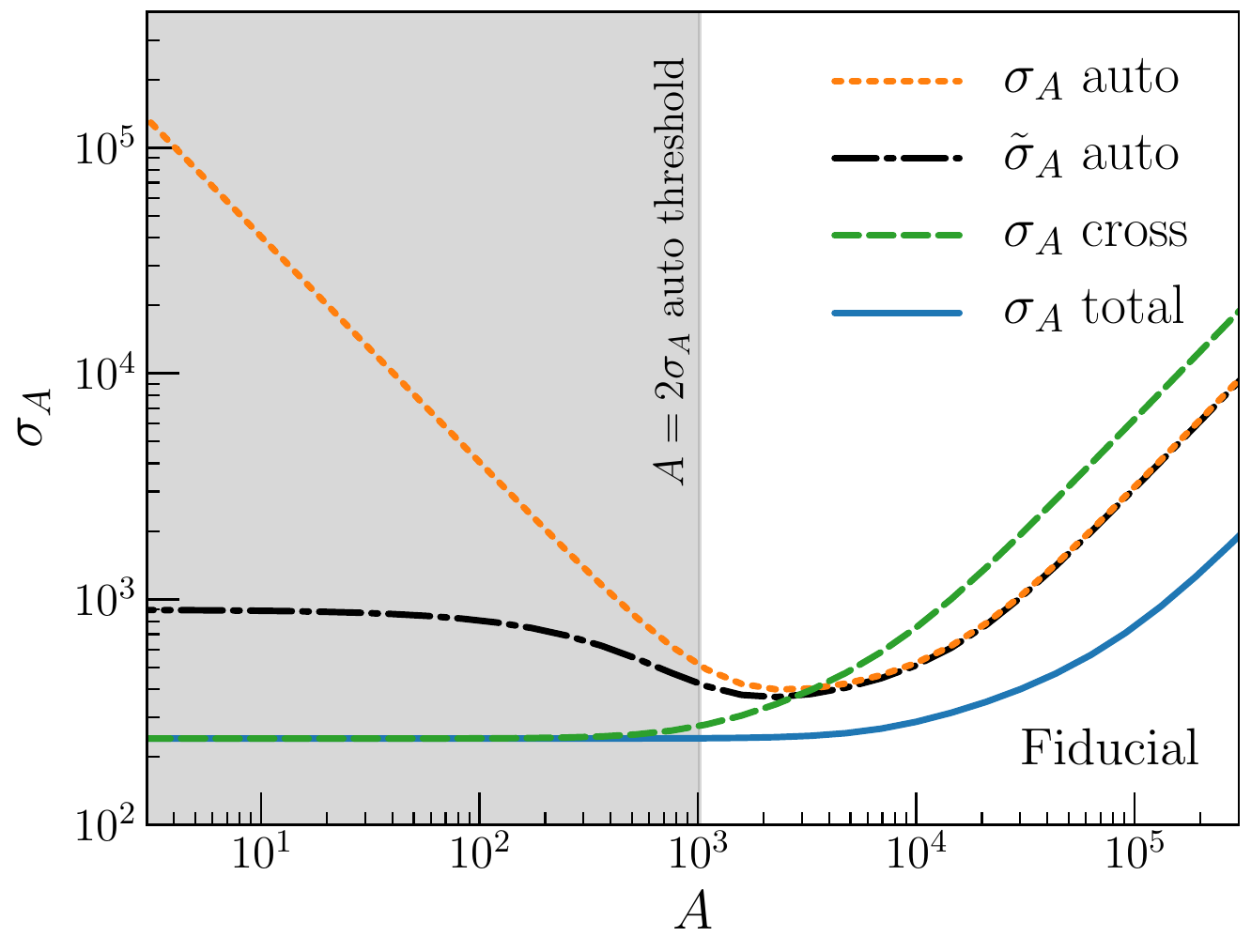}
\caption{Fisher error $\sigma_A$ as a function of the true value of the CIP amplitude $A$ using the CIP auto- ($\Delta\Delta$) and cross-($g\Delta$) spectrum, and their combination for the fiducial survey. Also shown is $\tilde{\sigma}_A$ (Eq.~\ref{fig:tilde_sigma}), the upper limit on $A$ implied by an upper limit on $A^2$. The two auto-spectrum results $\sigma_{A}$ and $\tilde{\sigma}_A$ agree well for $A \gtrsim 2\sigma_A$, which defines the regime where the Fisher results can be interpreted meaningfully. The cross-spectrum increasingly dominates the total result at low $A/\sigma_A$ while also contributing at high signal-to-noise ratio by reducing sample variance.}
\label{fig:detection_threshold}
\end{figure}

We show in Fig.~\ref{fig:detection_threshold} the results from the CIP auto-spectrum alone, the cross-spectrum alone, and their combination. As was the case with CIP-cross-CMB anisotropies~\cite{He:2015msa}, the cross-spectrum with galaxies here entirely dominates the total result at low signal-to-noise ratio $A/\sigma_A$. For large signal-to-noise the inclusion of the cross-spectrum helps to eliminate sample variance from the underlying random Gaussian field $\zeta$. 

For small CIP amplitude, the constraint from the auto-spectra alone has a subtlety \cite{He:2015msa}.
Since $P_{\Delta\Delta} \propto A ^2$, the Fisher error $\sigma_A \propto A^{-1}$ diverges as $A\rightarrow 0$ whereas the Fisher error on $A^2$ does not, $\sigma_{A^2} = 2|A| \sigma_{A}$. Of course an upper limit on $A^2$ would imply an upper limit on $A$
\beq 
\tilde{\sigma}_A = \sqrt{A^2 + \sigma_{A^2}} - |A|.
\label{fig:tilde_sigma}
\eeq
We show in Fig.~\ref{fig:detection_threshold} (following Ref.~\cite{He:2015msa}) that the two auto-spectrum errors $\sigma_A$ (orange dashed) and $\tilde{\sigma}_{A}$ (black dot-dashed) agree well for $A \gtrsim 2\sigma_{A}$ (unshaded region). Because in this regime we can meaningfully compare the Fisher errors, we choose to report as our main result the $2\sigma$ detection threshold. 

The behavior of the amplitude error and impact of including the cross-correlation can be understood analytically in the low and high signal-to-noise regime.\\

\textit{The low signal-to-noise regime -- } 
We can approximate the auto-spectrum error in the noise-dominated regime $P_{\Delta\Delta}(K) \ll N_{\Delta\Delta}(K)$ as
\beq
\sigma_A^{-2}\Big|_{\Delta\Delta} = \sum_K \frac{2 N_{\rm modes}}{A^2} \left( \frac{P_{\Delta\Delta}(K)}{N_{\Delta\Delta}(K)} \right)^2
\eeq
and the cross-spectrum errors as
\beq
\sigma_A^{-2}\Big|_{g\Delta} = \sum_K \frac{N_{\rm modes}}{A^2} \frac{P_{\Delta\Delta}(K)}{N_{\Delta\Delta}(K)} \frac{P_{gg}}{P_{gg} + \frac{1}{\bar{n}}} R_{g\Delta}^2,
\eeq
where 
\beq
N_{\rm modes} = V_{\rm survey} 4\pi K^2 dK/(2\pi)^3.
\eeq

Let us estimate that the $A = 2\sigma_A$ detection threshold occurs for the auto-spectrum at 
\beq
\frac{P_{\Delta\Delta}(K)}{N_{\Delta\Delta}(K)} \sim \left( \frac{K_{\rm min}}{K_{\Delta}}\right)^{3/2} ,
\eeq
where $K_{\Delta}$ be the representative wavenumber when the Fisher sum reaches half its total value. Then, at this level of signal, the constraint from the cross-spectrum alone is better than that from the auto-spectrum alone by a factor of 
\beq
\frac{\sigma_{g\Delta}^{-2}}{\sigma_{\Delta\Delta}^{-2}} \sim \frac{1}{1+(\bar{n}P)^{-1}}  \left (\frac{K_{\Delta}}{K_{\rm min}} \right) ^{3/2} R_{g\Delta}^2
\label{eq:improvement}
\eeq
where the signal correlation
\beq
R_{g\Delta}\equiv P_{g\Delta}/\sqrt{(P_{gg}P_{\Delta\Delta})} = 1
\eeq
drops out. 
 
Similarly to the CMB, the improvement from the cross-spectrum only depends on a few survey or analysis choices. As can be seen in Eq.~\ref{eq:improvement}, the improvement is larger for a survey with smaller shot noise. In addition, a larger cell volume means that a smaller range of CIP wavenumbers is accessible, so $K_{\Delta}$ is lower, and the improvement is smaller. For $r_{\rm cell, fid} \approx 3\, r_{\rm BAO}$ used for the fiducial survey in Fig.~\ref{fig:vsurvey} (we don't expect to use much smaller cell size than this), the improvement is about a factor of 2$-$3.
Finally, the ratio in Eq.~\ref{eq:improvement} scales as $V_{\rm survey}^{1/2}$ (or $f_{\rm sky}^{1/2}$ for CMB) -- the total number of modes available in the survey. So a survey with a larger volume would see a larger improvement, which is also reflected in Fig.~\ref{fig:vsurvey}.\\

\textit{The high signal-to-noise regime -- } In the high signal-to-noise regime, we can eliminate sample variance by jointly analyzing the auto- and the cross-spectra, taking their covariance into account. The high level of correlation between the various spectra is what makes the total results of Fig.~\ref{fig:detection_threshold} do much better at high $A$ than would a simple inverse squared sum. 
Interestingly, this effect seems to be more pronounced for the galaxy probes than for the CMB (see Fig.~6 of Ref.~\cite{He:2015msa}), possibly because in absence of shot noise the galaxy density $\delta_g$ is fully correlated with the adiabatic perturbations $\zeta$ as well as with the CIP fluctuations $R_{g\Delta} = 1$ (at least on the large scales that we consider), whereas the CMB anisotropies are not fully correlated with $\zeta$ nor with $\Delta$ because of the different projection effects from a 3-dimensional field to the last-scattering surface. The cross-correlation coefficient between $\Delta$ and CMB fields $Y \in \{T, E\}$
\beq
R_L^{Y\Delta} \equiv 
C_L^{Y\Delta}/\sqrt{(C_{L}^{YY} C_{L}^{\Delta\Delta})}
\eeq
oscillates between $-1$ and 1 as a function of $L$ (see Fig.~1 of Ref.~\cite{He:2015msa}), and as a result, averages to a factor less than one when averaged over $L$. However, even in the fully correlated case for the BAO, we still do not get infinite information as the result is still limited by the finite $\bar{n}P$ at high $A$ and by the large CIP measurement noise at low $A$.

\section{Discussion}
\label{sec:discussion}

\subsection{Comments on $H_0$ tension}
If CIPs exist, they can bias our current cosmological measurements that are made assuming no CIPs. In particular, given a BAO survey, CIPs on scales larger than the survey act as a super-sample fluctuation of the baryon density that could lead us to measure a different BAO scale than expected on average. A natural question to ask is: How much could our current BAO results be affected given the CIPs allowed by current data?

Recently, the \textit{Planck} team~\cite{Akrami:2018odb} used the smoothing of CMB peaks and the lensing potential power spectra to obtain $\Delta_{\rm rms}^2 = 0.0037^{+0.0016}_{-0.0021}$ for scale-invariant CIPs\footnote{These are not correlated with adiabatic perturbations as in the curvaton case, but their constraint can still be applied to correlated CIPs (see Appendix~\ref{sec:conversion}).}, ruling out 
$\Delta_{\rm rms}^2 \gtrsim 0.0069$ at 95\% confidence level. 
Since CMB measurements are affected by CIP fluctuations on the surface of last scattering (SLS), this measured $\Delta_{\rm rms, CMB}^2 = \langle |\Delta(\hat{n})|^2 \rangle$ receives contributions from scales above the sound horizon up to the size of the SLS. 

Now only CIPs on scales larger than the volume of the galaxy survey would contribute to the supersample effect. Let us assume that the CMB-measured r.m.s. fluctuations are similar to those inside the 3D sphere enclosed by the SLS $\Delta_{\rm rms, CMB}^2 \approx \Delta_{\rm rms, 3D}^2$. This assumption is reasonable since they share the same range of scales
\beq
\Delta_{\rm rms, 3D}^2 \propto \int_{k_{\rm min}}^{k_{\rm max}} k^2 P_{\Delta\Delta} dk 
\sim  \mathrm{log}\left(\frac{k_{\rm max}}{k_{\rm min}}\right),
\eeq
where $k_{\rm min}$ and  $k_{\rm max}$ are set by the size of the SLS and the BAO scale respectively, and where in the last step we used a scale-invariant CIP power spectrum 
$P_{\Delta\Delta}(k)\propto~k^{-3}$, which is approximate for the curvaton CIPs.
It turns out that for BOSS or any next generation galaxy survey, the survey volume would be roughly at the logarithmic middle between $k_{\rm min}$ and $k_{\rm max}$, so that about half the contribution to $\Delta_{\rm rms, 3D}^2$ would be from scales smaller than the survey volume, and about half for larger scales.

So taking half of the maximal value allowed by the Planck 95\% bound $\Delta_{\rm rms}^2 = \frac{1}{2}(0.0069)$ for the supersample effect, and assuming a $1\sigma$ fluctuation in the field $\Delta(\hat{\vx}) \approx \Delta_{\rm rms} = 0.059$, we get a bias on the inferred BAO scale $\tilde{\alpha}$ of $|\Delta/C| \approx 1.1\%$. 
Given that current experiments such as BOSS~\cite{Alam:2016hwk} have errors at the level of $1\%$, this corresponds to a bias at the $1\sigma$ level. For a $2\sigma$ fluctuation $\Delta(\hat{\vx}) \approx 2\Delta_{\rm rms}$, we would have a $2.2\%$ bias in $\alpha$. The current inverse distance ladder measurement is $H_0 = (67.9 \pm 0.8)~ \mathrm{km\, s^{-1}Mpc^{-1}}$~\cite{Aghanim:2018eyx}, in tension with the local measurement $H_0 = (73.52\pm1.62)\mathrm{\, km\, s^{-1}Mpc^{-1}}$~\cite{Riess:2018byc} at 3.1$\sigma$. A 2.2\% change in $H$ at $z = 0$ corresponds to roughly $\Delta H_0 \approx 1.5~\mathrm{km\, s^{-1}Mpc^{-1}}$, reducing the tension to 2.3$\sigma$. Note that although the 2$\sigma$ CIP maximally allowed by \textit{Planck} at 95\% confidence level could partially reduce the inverse ladder tension, it cannot resolve the tension between local and CMB-only $H_0$ measurements.

Finally, we caution the interpretation of CIPs as solving the \textit{Planck} lensing anomaly and possibly (partially) the $H_0$ tension since there is not yet a known physical mechanism to produce CIPs as high as that allowed by \textit{Planck} ($\Delta_{\rm rms}^2 = 0.0037$ is about $A^2 \approx (600)^2$). For comparison, the largest possible CIPs in the curvaton model is $A = 16$, giving a fractional bias in the BAO scale at the $\sim 3\times 10^{-4}$ level for a $1\sigma$ fluctuation, which translates to only $\sim0.1\%$ bias in $H_0$ in the isotropic case.\\

\subsection{Measuring $\delta_m^L$ using BAO Modulation} The dilation due to long-wavelength matter fluctuations $\delta_m^L$, which we considered as a potential bias of the CIP measurement from BAO, can also be used to measure the long-wavelength dark matter density from the modulation of the BAO scale that it induces --- similarly to the CIP estimator from BAO.
Since $r_{\rm BAO}=(1-\delta_m^L/3)\, \bar r_{\rm BAO}$, the r.m.s.~fluctuation of the BAO scale is $\langle|\delta_m^L|^2\rangle^{1/2}/3\simeq 2.5\% D(z)/D(0)$ if the r.m.s.~of the long mode is evaluated using the variance at the BAO scale \cite{2012PhRvD..85j3523S}.

A BAO modulation at this level is likely too small to be measurable with a survey like BOSS, because the BAO scale in the entire survey is only measured at the $1\%$ level.
As a consequence, the super-sample variance of the BAO scale from the long-mode dilation is also relatively small for a BOSS-like survey \cite{1711.00018}.

However, future surveys that probe larger volumes may be sensitive to a BAO modulation at this level, especially when cross-correlating the measured BAO modulation with the long mode measured directly from the survey galaxy density. 
This could be useful to estimate the matter power spectrum on very large scales.
Such a measurement could in principle be useful to constrain local primordial non-Gaussianity by avoiding potential observational systematics that can affect standard measurements of the large-scale power spectrum, such as star contamination (stars in the Galaxy that are mistakenly identified as extra-galactic galaxies, adding power on large angular scales corresponding to the angular extent of the Galaxy), or catastrophic redshift errors (low-redshift galaxies that are mistakenly attributed to high redshift, adding power on large angular scales).
While potentially reducing the impact of systematics, we expect this method to have larger statistical uncertainty than standard measurements of the large-scale power spectrum because the long mode estimated from BAO modulation is rather noisy.
\\

\section{Summary}
\label{sec:summary}

In this paper, we explored measuring CIPs from large-scale structure using the spatial modulation of the BAO scale that they induce:
As CIPs modulate the baryon-to-photon ratio spatially, the sound speed of acoustic waves before recombination is modulated, leading to a modulation of the BAO scale as a function of position on the sky. 

We forecasted the detectability of CIPs assuming a simple but non-optimal estimator: first divide the survey volume into many cells, then measure the local BAO scale deviation in each cell, and make a CIP map assuming a linear relationship. In the forecast, we assumed that the measurement noise is uncorrelated between cells, and that the other effects known to bias the BAO scale can be removed (e.g.~dilation effects from long modes $\delta_m^L$), either via BAO reconstruction, or via an alternative removal procedure described in Appendix~\ref{sec:bias_removal}.

Using the Fisher matrix formalism, we first compute the BAO error per cell, then the resulting noise of the  CIP modes measured from the modulation of the BAO across the cells, and from this the forecasted error of the CIP amplitude $A$ in the curvaton model using the reconstructed CIP and galaxy density auto- and cross-spectra as observables. 
For a fiducial survey similar to next-generation galaxy surveys like Euclid with $V_{\rm survey} = 27.8\, (\hGpc)^3$ obtained from $z = 0.75$ to 1.25, $f_{\rm sky} = 0.5$, $\bar{n}P = 5$, and $N_{\rm cell} = 1000$, we find a $2\sigma$ CIP detection threshold of 484, which is similar to WMAP constraints from the auto-correlation alone.

We explored how this result depends on various survey aspects (fixing all other parameters at the fiducial choices except specified otherwise). Varying the size of the cells, we find that there is a sweet spot for the cell size at $r_{\rm cell} \approx 2-3r_{\rm BAO}$.
For smaller cells, the BAO S/N per cell becomes too small, whereas for larger cells one cannot fit enough cells in the survey volume to observe the BAO modulation across cells. 

Given the fiducial survey volume, most of the fiducial settings are already close to optimal. 
When varying the survey volume and using a cosmic variance limited (CVL) survey, we find that it
 would be competitive with a stage 3 CMB experiment if one used emission line galaxies up to $z_{\rm max} \sim 7$, giving $2\sigma_A = 30$. These results are compatible with what we would expect from a naive mode counting estimate. If a larger volume is used, say up to $z \sim 150$, then the BAO method in its current form would start to become similar to the CMB CVL. Of course, there are no galaxies up to those high redshifts, so another tracer must be found to fully take advantage of the information in this large volume.

There is hope that 21cm experiments could -- in addition to offering BAO measurements at low redshift $z < 7$ -- offer a window into the $z>7$ range where it would be hard to observe galaxies. These 21cm measurements could extend galaxy BAO measurements to higher redshifts, as well as serve as potential tracers of the adiabatic perturbations there, to allow cross-correlations even at high redshifts. 

We found that using cross-correlations in the fiducial galaxy survey with $r_{\rm cell} \approx 3 \, r_{\rm BAO}$ already improved results by about a factor of 2.1 (from $2\sigma$ = 1023 to 484). 
We expect an improvement by a factor of about 2$-$3 for large future surveys, given the range of survey volume and cell size we explored, as the improvement decreases with cell size and increases with survey volume. 

As in the CMB case, the ultimate limit to the method is the BAO scale. For CIPs much smaller than the sound horizon, the sound waves propagate at a speed that's modulated many times before recombination, so the net effect is too small to be observable. Therefore, the CMB and BAO methods presented here are ultimately limited by the same pre-recombination physics. However, they will see different systematics challenges from post-recombination processes. 

On the BAO side, accurate measurements require an
accurate BAO reconstruction and knowing the right galaxy bias in order to infer the large-scale $\delta_m^L$ from $\delta_g^L$ and correct for the effects of $\delta_m^L$.
Because the BAO and CMB suffer from different systematics, they can potentially be used as a cross-check for each other and/or be used in a joint analysis. 
For example, if CIPs are detected in the BAOs but not in the CMB, it could point us to new physics that modulate the BAO scale or to systematics in the BAO survey. On the other hand, if CIPs are detected in the CMB -- say from the extra smoothing of peaks in the \textit{Planck} CMB power spectra which is larger than lensing can account for (see the lensing anomaly~\cite{Akrami:2018odb}) -- but they are not detected in the BAOs, then we may rule out CIPs as a potential cause of the anomaly and look for other effects.

The presence of CIPs -- at least at the level allowed by the \textit{Planck} power spectra measurement -- could bias our current measurements of the BAO scale. For BOSS or any next generation survey, a 1$\sigma$ (2$\sigma$) fluctuation can cause a 1.1\% (2.2\%) bias in $\alpha$, leading to a similar bias in the $H_0$ measurement. This type of supersample fluctuation could partially account for the current tension between the inverse-distance ladder and local measurement of $H_0$. However, we expect the amount of possible bias to reduce as future constraints on CIPs become tighter, as we do not yet know a physical mechanism that could give rise to the high amount of CIPs corresponding to the maximal value allowed by current \textit{Planck} 95\% constraints.

Finally, we note that unlike the CMB forecast we compare to, the simple method of measuring BAO scale modulation on the sky is not optimized for measuring CIPs. To do so, a minimum-variance estimator $\hat{\Delta}^{\rm mv}(K)$ should be built by optimally weighing the correlations between the different wavenumber pairs $k, k' \gg K$ of galaxy densities.
This means that there is still room for improvement on the sensitivity reported here that one can explore in future work. 
 Another way to improve upon the current forecast is to use multiple tracers of the adiabatic perturbations to cancel sample variance, for example by using different types of galaxies. \\

In conclusion, searching for a spatial modulation of the BAO scale represents a new tool to use the BAO feature as a window into primordial physics, specifically to constrain compensated isocurvature perturbations that modulate the sound speed of the baryon-photon fluid. This adds to the previous applications of using the BAO as a standard ruler \cite{1998ApJ...504L..57E} and as a tool to search for light particles \cite{Baumann:2017gkg,2018arXiv180310741B}. It also adds to and opens up new possibilities for synergies with other CIP probes such as CMB power spectra and quadratic estimators, luminosity-weighted galaxy correlation functions and 21cm measurements.

\begin{acknowledgements}
We thank Olivier Dor\'e, Marko Simonovi\'c and Matias Zaldarriaga for instructive discussions. We are very grateful to Olivier Dor\'e, Daniel Grin, Marko Simonovi\'c and Tristan Smith for feedback on the manuscript, as well as Tzu-Ching Chang and Daniel Lenz for other useful feedback. 
We thank Zvonimir Vlah for providing the nowiggle power spectrum.
Part of this work was done at Jet Propulsion Laboratory, California Institute of Technology, under a contract with the National Aeronautics and Space Administration. MS gratefully acknowledges support from the Jeff Bezos and Corning Glass Works Fellowships at IAS as well as the National Science Foundation. 
\end{acknowledgements}

\appendix

\section{Explicit Removal of the Dilation Bias}
\label{sec:bias_removal}
We have argued in the main text that applying BAO reconstruction globally before dividing the survey into cells takes care of the potential biases of the CIP measurement due to the  BAO modulation induced by dilation and bulk flows.
In this appendix we present an alternative
procedure where we treat separately the dilation effect induced by very long modes, i.e.~modes longer than the cell size.
This procedure works as follows.

\begin{enumerate}
\item{Divide the total survey volume $V_{\rm tot}$ into $N_{\rm cell}$ cells with the same volume $V_{\rm cell}$ centered at positions $\VX_i$, where $i = 1 ... N_{\rm cell}$.}

\item{Measure the BAO scale $\hat{\tilde{\alpha}}(\VX_i)$ using galaxies in each cell $\VX_i$, where $\tilde{\alpha}$ is defined with respect to fiducial cosmology measured by \textit{Planck}.}

\item{Perform BAO reconstruction locally in each cell, using only large-scale modes within each cell.
} 

\item{Measure the galaxy density $\delta_{g}^{L}(\VK)$ on scales larger than the cell, but smaller than the survey size. Infer the long-wavelength matter density modes using $\hat{\delta}_{m}^{L}(\VK) \approx \hat{\delta}_{g}^{L}(\VK)/b(z)$ where $b(z)$ is the linear galaxy bias at redshift $z$, and smoothing on the scale of the cell size is applied. 
}
 
\item{Obtain the CIP estimator using the Fourier transform of the BAO map $\hat{\tilde{\alpha}}(\VK)$, and remove the dilation bias due to $\delta_m^L(\VK)$ measured in step 4
\beq
\hat{\Delta}(\VK)  = C \left( \hat{\tilde{\alpha}}(\VK) - 1\right) + \frac{C \hat{\delta}_m^L(\VK)}{3}.
\eeq
}
\item{Compute the CIP auto-spectrum $\hat{P}_{\Delta\Delta}(K)$ and cross-spectrum $\hat{P}_{g\Delta}(K)$ with the long-wavelength galaxy density $\delta_{g}^{L}(\VK)$, and use them to measure the CIP amplitude $A$ in the curvaton model.} 
\end{enumerate}

In step 3, the BAO reconstruction uses only modes inside each cell. This is sufficient to reduce most of the non-linear peak smearing because density modes larger than the BAO scale or cell size move two galaxies separated by $r_{\rm BAO}$ by the same amount, so they do not shift the BAO scale.
In fact, reconstructing only modes on scales close to the BAO scale should be sufficient because the non-linear peak smearing, given by
\cite{Baldauf:2015xfa}
\beq
\Sigma^2 = \frac{1}{3}\int \frac{dq}{2\pi^2} [1-j_0(qr_{\rm BAO}) + 2j_2(qr_{\rm BAO})] P(q),
\eeq
where $j_n$ is the $n^{th}$ order spherical Bessel function, is dominated by bulk flows on scales around the BAO scale: for $q \ll 1/r_{\rm BAO}$ the term in square brackets is suppressed, whereas for $q \gg 1/r_{\rm BAO}$ the power spectrum term becomes suppressed. 

Note that contributions to $\Sigma$ only impact the uncertainty of the measured BAO scale without leading to a bias.
There is an effect, however, that shifts the BAO scale from cell to cell that we propose to mitigate using steps 4 and 5. A long-wavelength mode $\delta_m^L(\VK)$ varying on scales larger than the cell size (the cells are assumed to be larger than the BAO scale), have effects indistinguishable (at first order in $\delta_m^L$) to a separate universe with local curvature. In this locally curved universe, small scales inside the cell are contracted by a factor of 
\beq
\frac{a_\mathrm{curved}(\delta_m^L)}{a_0} = \left(1+\delta_m^L\right)^{-1/3} \approx \left(1-\frac{\delta_m^L}{3}\right), 
\eeq
shifting the BAO scale in linear theory as \cite{2012PhRvD..85j3523S}
\beq
r_{\rm{BAO}} \rightarrow r_{\rm{BAO}}\left(1-\frac{\delta_m^L}{3}\right).
\eeq

In Ref.~\cite{2012PhRvD..85j3523S}, a second effect biasing the global BAO scale measurement was considered: The enhanced growth of local perturbations in a closed universe
and the conversion between local and global mean densities
resulting in an enhanced contribution from regions of positive $\delta_m$ to the global correlation function.
We argue, however, that this effect is not relevant for us because 
we do not combine the measurements in different cells to give one global BAO scale measurement. 
Furthermore, for $\delta_m^L$ on scales smaller than the cell size, we assume that the local BAO reconstruction would revert its effect. 

Therefore, in the presence of large-scale CIPs $\Delta(\VX_i)$ and matter density fluctuations $\delta_m^L(\VX_i)$, the local BAO scale at $\VX_i$ is shifted as

\bea
r_{\rm BAO}(\VX_i) &\approx& r_{\rm BAO}^{0} \left(1 - \frac{\delta_m^L(\VX_i)}{3}\right) \left(1 + \frac{\Delta(\VX_i)}{C}\right) \notag\\
& \approx &r_{\rm BAO}^{0} \left(1 - \frac{\delta_m^L(\VX_i)}{3}  + \frac{\Delta(\VX_i)}{C}\right) \notag\\
 & = &r_{\rm BAO}^{0} \left(1 + \frac{\Delta_{\rm wrong}(\VX_i)}{C}\right),
\eea
where $r_{\rm BAO}^{0}$ is the unmodulated BAO scale in absence of $\Delta(\VX_i)$ and $\delta_m^L(\VX_i)$. 

If uncorrected, the dilation effect from a positive $\delta_m(\VX_i)$ fluctuation would therefore contribute to a negative bias of the CIP measurement
\beq
\Delta_{\rm wrong}(\VX_i) = \Delta(\VX_i) - C \delta_m^L(\VX_i)/3,
\eeq
or correspondingly in Fourier space (when the linear approximation holds)
\beq
\Delta_{\rm wrong}(\VK) = \Delta(\VK) - C \delta_m^L(\VK)/3.
\eeq
Note that the above holds for CIP scales larger than the cell size but smaller than the survey size $V_{\rm cell}^{1/3} \lesssim 2\pi/K \lesssim V_{\rm survey}^{1/3}$, for we cannot measure $\Delta$ nor $\delta_m^L$ on scales larger than the survey, and for scales smaller than the cell size, we assume that the BAO reconstruction can mitigate the bias.

In step 4, we measure $\hat{\delta}_m^L(\VX_i) = \hat{\delta}_g^{L}(\VX_i)/b(z)$ and use it in step 5 to correct the bias. 
Since the correction is realization-dependent, there is no cosmic variance that would contribute to a noise bias in the CIP power spectra. 
However, modeling uncertainties in $\delta_m^L(\VX_i)$ and $b(z)$ could lead to a residual bias, but we expect this to be small given that the relevant scales are larger than the BAO scale, and linear theory holds well on these scales. 

Finally, one might worry about the dilation bias from $\delta_m^L(\VX_i)$ inside the cell when the BAO reconstruction is not perfect. We argue that since the bias would fluctuate many times inside the cell, most of its contributions cancel as the BAO scale measurement is an averaged quantity of the cell (just like the effect of CIP modes with many wavelengths inside the sound horizon).  \\

\section{CIP Conventions and Conversion Rules} 
\label{sec:conversion}
There are a few different conventions used in the literature for scale-invariant CIPs. Here we briefly review these conventions and provide approximate rules to convert between these conventions and the correlated CIP amplitude $A$ that we use in this paper. 

CMB measurements probe the CIP field on the surface of last scattering
\begin{equation}
\Delta(\hat n) = \Delta(\vx = \chi_* \hat n), 
\end{equation}
where $\chi_*$ is the comoving distance to the SLS.
Assuming instantaneous recombination for the field $\Delta$, its angular power spectrum is
\beq
C_L^{\Delta \Delta} = \int k^2d k P_{\Delta \Delta}(k) j^2_L(k \chi_*). 
\eeq	
For correlated CIPs (e.g.~of curvaton origin) we used  in this paper 
\beq
P_{\Delta \Delta}^{\rm corr}(k) = A^2 P_{\zeta\zeta}, 
\eeq
whereas for the scale-invariant CIP power spectrum, a common convention is
\beq
P_{\Delta \Delta}^{\rm SI}(k) = A_{\rm CIP}k^{-3},
\eeq
in which case we can approximate
\begin{equation}
C_L^{\Delta \Delta, \rm SI} \approx \frac{A_{\rm CIP}}{\pi L(L+1)}.
\end{equation}

Note that the correlated CIPs are nearly scale-invariant because of the small tilt $n_s -1$. We derive approximate conversions rules below by comparing a small range of $C_L^{\Delta\Delta, \rm corr}$ at low $L$ to $C_L^{\Delta \Delta, \rm SI}$ such that for that range
\beq
\frac{A_{\rm CIP}}{\pi L(L+1)} \approx A^2 \left(\frac{C\DD}{A^2}\right).
\eeq
Taking the average over the range $2 \leq L \leq 37$ we get $\langle\left(C\DD/A^2\right)L(L+1)\rangle_{L \in [2, 37]} = 1.462\times10^{-8}$ and so
\beq
\label{eq:conversion_Acip}
A_{\rm CIP} \approx (4.594 \times 10^{-8}) A^2. 
\eeq

The \textit{Planck} constraints so far have been focused on using the smoothing of the acoustic peaks and the contamination to the lensing potential power spectrum that scale-invariant CIPs would induce. These contributions are expressed in terms of $\Delta_{\rm rms}(R)$ the root-mean-square CIP amplitude over some length-scale $R$:
\begin{equation}
\Delta^2_{\rm rms}(R) = \frac{1}{2\pi^2} \int k^2 dk[3 j_1(kR)/(kR)]^2 P_{\Delta \Delta}(k).
\label{eq:DeltaRMS}
\end{equation}
For the CMB calculations, the relevant scale is the size of the surface of last scattering $R_{\rm CMB}$
\begin{equation}
\Delta^2_{\rm rms}(R_{\rm CMB}) \equiv \langle \Delta^{2}(\hat n) \rangle  = \sum_{L=1}^{100} \frac{2L+1}{4\pi} C_L^{\Delta \Delta}.
\end{equation}
Numerically evaluating the sum, we get
\begin{equation}
\Delta^2_{\rm rms}(R_{\rm CMB}) \approx 0.2377 A_{\rm CIP},
\end{equation}	
for a scale-invariant power spectrum.

With the caveat that CIP contributions to the CMB and lensing power spectra expressed in terms of $\Delta_{\rm rms}$ may differ for scale-invariant and correlated CIPs, we can convert the central value of the latest \textit{Planck} constraint $\Delta_{\rm rms}^2 = 0.0037$ (or $A_{\rm {CIP}} \approx 0.015$) to roughly $A^2 \approx (580)^2$ 
\beq
\Delta_{\rm rms}(R_{\rm CMB}) \approx (1.045 \times 10^{-4} ) A.
\eeq

The WMAP constraints~\cite{Grin:2013uya} on the other hand used a quadratic CMB temperature estimator to put an upper limit on the scale-invariant CIPs amplitude $A_{\rm CIP}/\pi$ as well as the individual multipoles of $C_L^{\Delta\Delta}$ without any model assumption. Note that a different usage of $A$ was employed there, $A \equiv A_{\rm CIP}/\pi$ (as well as in Ref.~\cite{Grin:2011nk} where $A \equiv A_{\rm CIP}$.) These usages of $A$ are different from the notation in this paper and Refs.~\cite{He:2015msa, Heinrich:2016gqe, Heinrich:2017psm} to denote the amplitude of correlated CIPs.

Ref.~\cite{Grin:2013uya} also stated their constraints in terms of $\Delta_{\rm cl}$, to facilitate comparison with cluster constraints from Ref.~\cite{Holder:2009gd}. Here $\Delta_{\rm cl}\equiv \Delta^2_{\rm rms}(R_{\rm cl})$, where $R_{\rm cl}$ is the mean separation between galaxy clusters. Taking $k_{\rm min} \approx (10\, \mathrm{Gpc})^{-1}$ and $R_{\rm cl} \approx 10\, \mathrm{Mpc}$ in Eq.~\ref{eq:DeltaRMS}, we have 
\beq
\Delta_{\rm cl}^2 \approx \frac{A_{\rm CIP}\, \rm{ln}(1000)}{2\pi^2}. 
\eeq
Using Eq.~\ref{eq:conversion_Acip} we get
	\beq \label{eq:conversion_Deltacl}
	\Delta_{\rm cl}\approx(1.268\times10^{-4})A.
	\eeq

Using these conversion rules, the WMAP 95\% upper bound of $A_{\rm CIP}/\pi < 0.011$ (or $\Delta_{\rm cl}^2 < 0.012$) corresponds to $A^2 < (870)^2$ for correlated CIPs. Note that this is a $2\sigma$ upper bound in $A^2$: A $2\sigma$ detection in $A^2$ actually corresponds to a $4\sigma$ detection in $A$, so a $2\sigma$ detection in $A$ would occur at a smaller $A$, probably around $2\sigma_A \sim 500$. This is why we say that the  fiducial galaxy survey considered in this work with $2\sigma_A = 484$ is similar in sensitivity to the current WMAP constraint.

\bibliography{cipbao.bib}

\end{document}